\documentclass[pdflatex,sn-standardnature]{sn-jnl}

\usepackage{caption}
\usepackage{subcaption}
\usepackage{graphicx}

\jyear{2022}%

\theoremstyle{thmstyleone}%
\theoremstyle{thmstyletwo}%

\theoremstyle{thmstylethree}%

\begin{document}

\title[Temperature inhomogeneities in Mrk 71 can not be discarded]{Temperature inhomogeneities in Mrk 71 can not be discarded}


\author*[1]{\fnm{J. Eduardo} \sur{M\'endez-Delgado}}

\author[2,3]{\fnm{C\'esar} \sur{Esteban}}

\author[2,3]{\fnm{Jorge} \sur{Garc\'ia-Rojas}}

\author[1]{\fnm{Kathryn} \sur{Kreckel}}

\author[4]{\fnm{Manuel} \sur{Peimbert}}

\affil*[1]{\orgdiv{Astronomisches Rechen-Institut}, \orgname{Zentrum f\"ur Astronomie der Universit\"at Heidelberg}, \orgaddress{\street{M\"onchhofstraße 12-14}, \city{Heidelberg}, \postcode{D-69120}, \state{Baden-W\"urttemberg}, \country{Germany}}}

\affil[2]{\orgdiv{Instituto de Astrof\'isica de Canarias}, \orgname{(IAC)}, \orgaddress{\street{V\'ia L\'actea, 1}, \city{San Crist\'obal de La Laguna}, \postcode{E-38205}, \state{Santa Cruz de Tenerife}, \country{Spain}}}

\affil[3]{\orgdiv{Departamento de Astrof\'isica}, \orgname{Universidad de La Laguna}, \orgaddress{\street{Astrof\'isico Francisco S\'anchez, s/n.}, \city{San Crist\'obal de La Laguna}, \postcode{E-38206}, \state{Santa Cruz de Tenerife}, \country{Spain}}}

\affil[4]{\orgdiv{Instituto de Astronom\'ia}, \orgname{Universidad Nacional Aut\'onoma de M\'exico}, \orgaddress{\street{Apartado Postal 70-264}, \city{Coyoac\'an}, \postcode{04510}, \state{Mexico City}, \country{Mexico}}}

\maketitle

\noindent \textbf{In a very recent work, \cite{Chen:2023} claim that the scenario of temperature inhomogeneities proposed by \cite{Peimbert:1967} ($t^2>0$) is not able to explain the O$^{2+}$/H$^{+}$ abundance discrepancy observed between the calculations based on the optical [O\,III] collisional excited lines (CELs) and the O\,II recombination lines (RLs) in the star forming galaxy Mrk~71. In this work, we show that conclusions of \cite{Chen:2023} depend on several assumptions on the absolute flux calibration, reddening correction and the adopted electron density. In fact, using the data of \cite{Chen:2023} in a different way and even considering their 1$\sigma$ uncertainties, it is possible to reach the opposite conclusion, consistent with $t^2=0.097 ^{+0.008} _{-0.009}$. Therefore, the existence of temperature inhomogeneities causing the O$^{2+}$/H$^{+}$ abundance discrepancy in Mrk~71 can not be ruled out.}\\


\renewcommand\figurename{Figure}

\cite{Chen:2023} tested the presence of temperature inhomogeneities in the star forming galaxy Mrk~71. To carry out their analysis, \cite{Chen:2023} used an optical spectrum  from the Keck Cosmic Web Imager (KCWI) at the W. M. Keck Observatory as well as IR spectra from the Far Infrared Field-Imaging Line Spectrometer (FIFI-LS) at
the Stratospheric Observatory for Infrared Astronomy (SOFIA) and from the Photodetector Array
Camera and Spectrometer (PACS) at the Herschel Space Observatory. 


By comparing the different H\,I line flux ratios with the theoretical predictions, they infer a reddening constant $c(\text{H}\beta)=0.09\pm 0.04$, considering the reddening curve of \cite{Cardelli:1989} with $R_V=3.1$. They derive the electron density ($n_e$) of the gas with three indicators:  [O\,II] $\lambda 3726/\lambda 3729$, [O\,III] $\lambda 52\mu \text{m}/\lambda 88\mu \text{m}$ and the O\,II V1 RL multiplet. On the other hand, they derive the $T_e$ by considering [O\,III]  $\lambda 4363/\lambda 4959$ as well as [O\,III]  $\lambda 4959/\lambda 52\mu \text{m}$ and $\lambda 4959/\lambda 88\mu \text{m}$. From the comparison of the O$^{2+}$/H$^{+}$ abundance derived both with optical [O\,III] CELs and O\,II RLs and assuming that the abundance discrepancy (AD) is produced by temperature variations, they infer a $t^2\sim 0.1$ (see Eq.~(12) from \cite{Peimbert:1967}). This result would imply that both the derived temperature from [O\,III] $\lambda 4959/\lambda 52\mu \text{m}$ and $\lambda 4959/\lambda 88\mu \text{m}$ should be $\sim 3000 \text{ K}$ lower than what is obtained from [O\,III]  $\lambda 4363/\lambda 4959$. However, \cite{Chen:2023} found a good consistency between their calculations of $T_e$ based on [O\,III] $\lambda 4959/\lambda 52\mu \text{m}$,  $\lambda 4959/\lambda 88\mu \text{m}$ and $\lambda 4363/\lambda 4959$ and therefore they claim the absence of significant temperature fluctuations in Mrk~71.

The results obtained by \cite{Chen:2023} are highly dependent on the accuracy of the absolute flux calibration between the three instruments, since there are no H\,I detections in the IR data that could be used to normalize the spectra. Considering that the KCWI observations were taken under non-photometric conditions \cite{Chen:2023}, that the FIFI-LS observations present telluric features \cite{Sutter:2022} and that the PACS observations were carried out in the ``un-chopped'' mode and show detector response variations \cite{Fadda:2016}, the absolute flux calibration between the three different kinds of data is not straightforward. In fact, the comparison of [C\,II] $\lambda 158\mu \text{m}$, detected both in FIFI-LS and PACS reveals a difference of $\sim 15\%$ between the flux calibrated data of both instruments even after the PACS detector response variations correction. Possible systematic differences between the optical and IR-spectra are not analyzed or quantified by \cite{Chen:2023}.

The difference between the FIFI-LS and PACS spectra implies the existence of a systematic bias in the flux of at least one of the [O\,III] IR CELs used. Dividing the flux difference of $\sim 15\%$ quadratically and including it in the uncertainty bars does not properly treat the systematic error, as it impacts differently [O\,III] $\lambda 4959/\lambda 52\mu \text{m}$ than [O\,III] $\lambda 4959/\lambda 88\mu \text{m}$, given their different $n_{\rm e}$-dependence. Considering that both [O\,III] IR CELs have similar fluxes, a more robust way to reduce the impact of the flux bias is to use the sum of their fluxes instead, comparing $T_{\rm e}$ derived from [O\,III] $\lambda 4959/\lambda \lambda 52 + 88\mu \text{m}$ with the value obtained using $\lambda 4363/\lambda 4959$. The flux systematic difference also calls into question the density derived from [O\,III] $\lambda 52\mu \text{m}/\lambda 88\mu \text{m}$ with the reported absolute fluxes.

\begin{figure}
    
    \centering
    \begin{subfigure}[t]{0.49\textwidth}
        \centering
        \includegraphics[width=\linewidth]{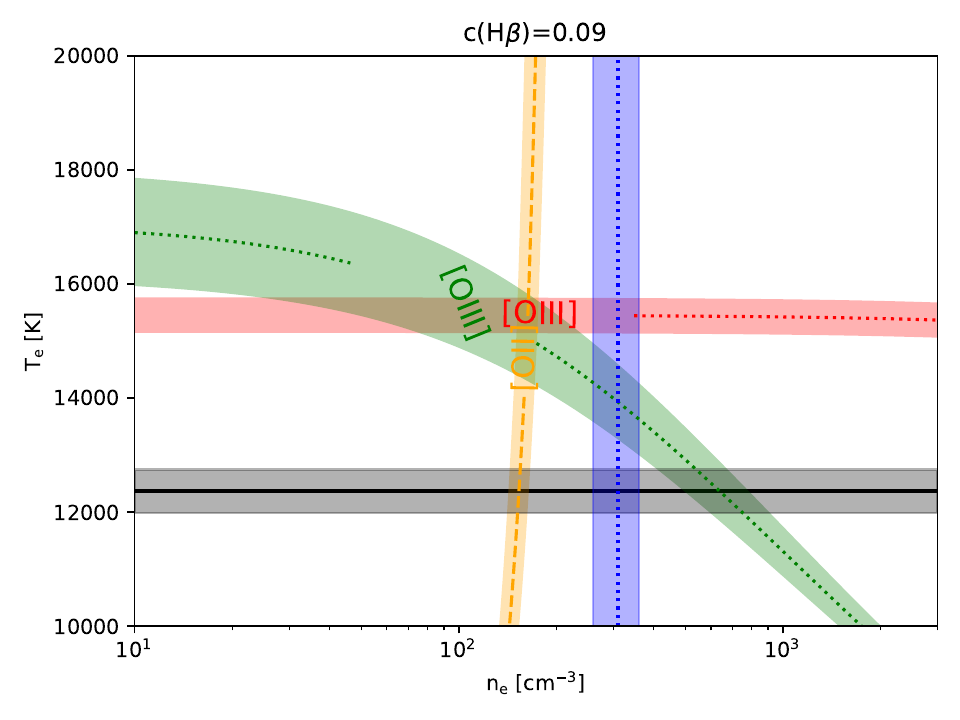} 
    \end{subfigure}
    \begin{subfigure}[t]{0.49\textwidth}
        \centering
        \includegraphics[width=\linewidth]{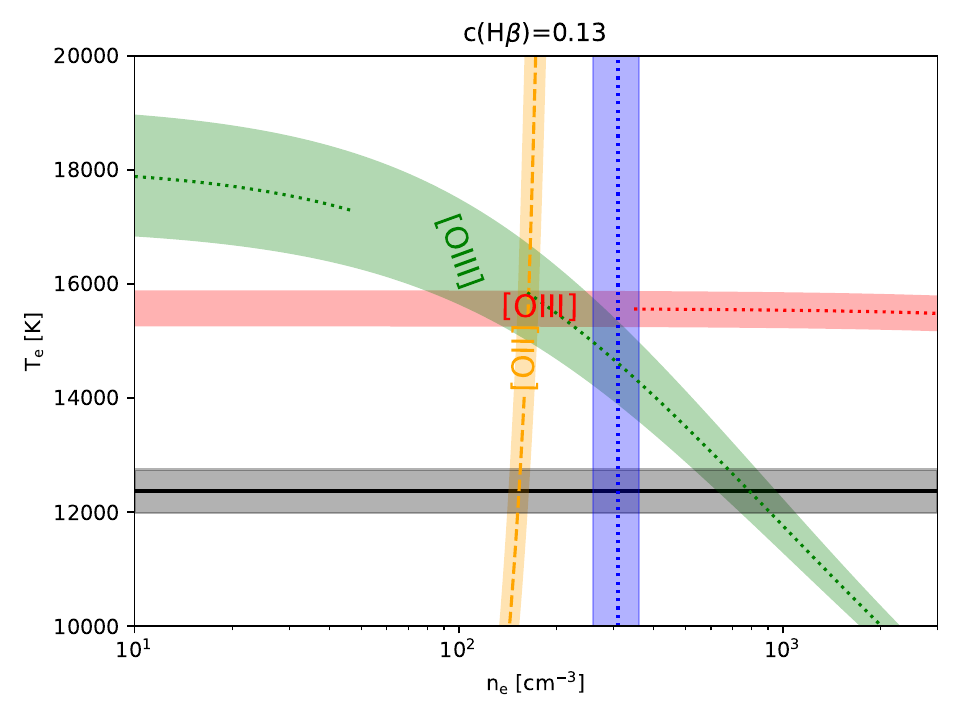} 
    \end{subfigure}

    \begin{subfigure}[t]{0.6\textwidth}
    \centering
        \includegraphics[width=\linewidth]{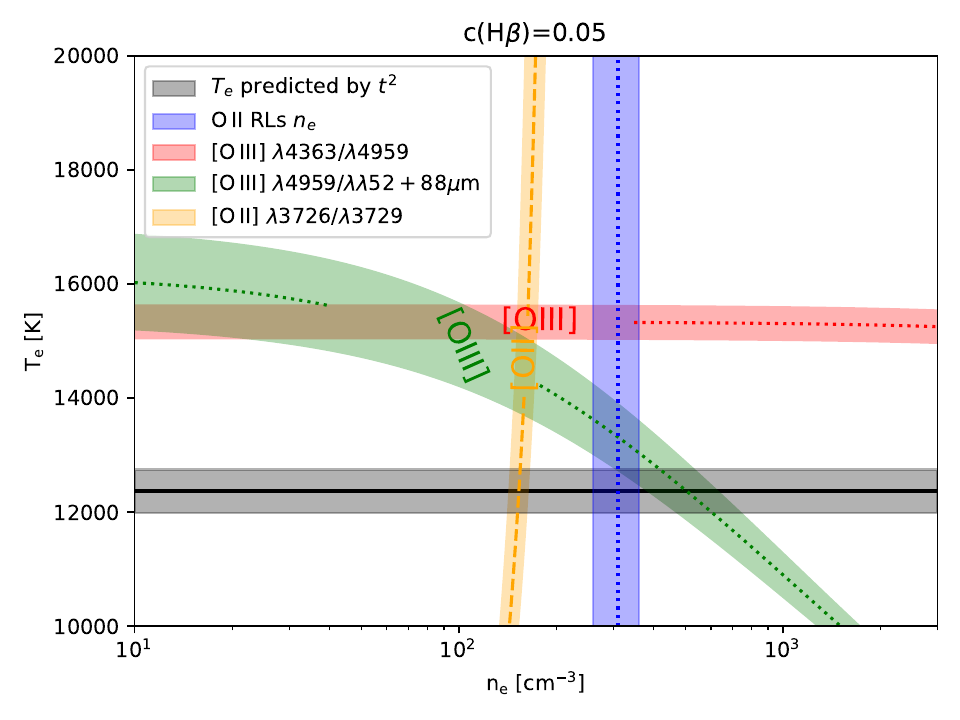} 
    \end{subfigure}
    \caption{\textbf{Using the data presented by \cite{Chen:2023} it is possible to show either the absence of temperature inhomogeneities or the opposite case.}  Each panel shows a plasma diagnostic plot considering a slightly different reddening constant, $c(\text{H}\beta)$, all consistent within the 1$\sigma$ uncertainties. According to the $t^2$-paradigm proposed by \cite{Peimbert:1967}, in the presence of temperature inhomogeneities, [O\,III] $\lambda 4959/\lambda \lambda 52 + 88\mu \text{m}$ (green band) should be lower than [O\,III] $\lambda 4363/\lambda 4959$ (red band), matching the value predicted by the O\,II RLs (black band). Depending on
    $c(\text{H}\beta)$ and the adopted electron density ($n_e$), it is possible to argue either against the existence of temperature fluctuations (as concluded by \cite{Chen:2023}) or the opposite case.}\label{fig:fig1}
\end{figure}

In Fig.~\ref{fig:fig1} we present the resulting plasma diagnostics considering the line fluxes and uncertainties reported by \cite{Chen:2023} under three values of $c(\text{H}\beta)$, all consistent within the reported 1$\sigma$. In all cases, the reddening curve of \cite{Cardelli:1989} with $R_V=3.1$ was used. The atomic data used were the default ones from PyNeb \cite{Luridiana:2015} in its version 1.1.16, the same ones assumed by \cite{Chen:2023}. As shown in Fig.~\ref{fig:fig1}, depending on the $c(\text{H}\beta)$ and the $n_e$ adopted, it is possible to be consistent either with the absence of temperature inhomogeneities (as \cite{Chen:2023} concluded) or the opposite case, predicted by $t^2=0.097^{+0.008} _{-0.009}$.

The adoption of a density value close to $n_e$([O\,II] $\lambda3726/\lambda3729$)=$160 \pm 10 \text{ cm}^{-3}$ instead of the available $n_e$(O\,II)=$310 \pm 50 \text{ cm}^{-3}$ was not justified by \cite{Chen:2023}. Both values could be considered typical within the range of densities found in previous studies of Mrk~71 \cite{Gonzalez-Delgado:1994,Esteban:2002,Esteban:2009}. The most evident problem with the $n_e$ value derived from [O\,II] $\lambda3726/\lambda 3729$ by \cite{Chen:2023} is that panels (c) and (d) of their Fig.~2 show that the spectral resolution of KCWI is insufficient to have at least a partial separation of the [O\,II] doublet. Therefore, this density value strongly depends on the assumptions imposed on the Gaussian deblend required to measure the lines separately with uncertainty bars of $\sim 0.6\%$.



In order to reinforce their arguments on the absence of strong density and temperature
fluctuations in Mrk~71, \cite{Chen:2023} mention that in a sample of regions, among which Mrk~71 does not appear, \cite{Mingozzi:2022} found differences of only $\sim 1000\text{ K}$ between [O\,III]  $\lambda 1666/\lambda 5007$ and [O\,III] and $\lambda 4363/\lambda 5007$ and that this is not sufficient for the temperature fluctuations scenario to explain the observed AD factor. However, such an assertion can be proven to be incorrect. Considering Eq.~15 from \cite{Peimbert:1967}: $T_e(\text{[O\,III]} \lambda 1666/\lambda 5007 )-T_e(\text{[O\,III]} \lambda 4363/\lambda 5007 ) [\text{K}]=12330 \times t^2$. For the most common value of $t^2\sim0.04$ found for the star forming regions \cite{garcia_rojas:2007,Peimbert:2012}, the temperature difference would be of $\sim 500\text{ K}$. A very extreme value of $t^2=0.097 ^{+0.008} _{-0.009}$, would imply a difference of $1200 \pm 110\text{ K}$. This exercise demonstrates that the temperature differences found by \cite{Mingozzi:2022} can comprise typical and extreme values of $t^2$.




We conclude that based on the data and the analysis presented by \cite{Chen:2023}, one cannot be conclusive on the presence or absence of temperature inhomogeneities in Mrk~71, since both interpretations are possible even within their estimated 1$\sigma$ uncertainties. To be conclusive in this regard, it is necessary to consider the possible presence of density variations that could introduce systematic biases on the $n_e$ diagnostics even if the line intensity ratios are well measured \cite{Rubin:1989}. $n_e$-biases could affect determinations of O$^{2+}$/H$^{+}$ based on [O\,III] $\lambda \lambda 52+88\mu\text{m}$ in a much higher extent than those based on O\,II V1. Observational evidence of temperature and density inhomogeneities in star-forming regions (including Mrk~71) is presented by \cite{mendez2023a} and \cite{mendez2023b}, respectively.

\clearpage


\backmatter

\bmhead{Acknowledgments}
JEM-D thanks the help provided by V. Gomez-Llanos in managing the assignment of colors in the PyNeb plasma diagnostics and to O. V. Egorov for fruitful discussions.

\bmhead{Authors' contributions}
JEM-D lead the analysis and writing of the manuscript. CE, JG-R, KK and MP provided critical feedback and modified the text.

\bmhead{Conflict of interest/Competing interests}

The authors declare that they have no competing financial interests.

\bmhead{Data availability}

All the data discussed here was presented by \cite{Chen:2023}.

\bmhead{Code availability}
Our results use the PyNeb code,  publicly available on GitHub.
https://github.com/Morisset/PyNeb\_devel

\bmhead{Funding}

JEM-D and KK gratefully acknowledge funding from the Deutsche Forschungsgemeinschaft (DFG, German Research Foundation) in the form of an Emmy Noether Research Group (grant number KR4598/2-1, PI Kreckel).  CE and JG-R acknowledge support from the Agencia Estatal de Investigaci\'on del Ministerio de Ciencia e Innovaci\'on (AEI-MCINN) under grant {\it Espectroscop\'ia de campo integral de regiones H\,II locales. Modelos para el estudio de regiones H\,II extragal\'acticas} with reference 10.13039/501100011033 and support under grant P/308614 financed by funds transferred from the Spanish Ministry of Science, Innovation and Universities, charged to the General State Budgets and with funds transferred from the General Budgets of the Autonomous Community of the Canary Islands by the MCIU. JG-R acknowledges support from an Advanced Fellowship under the Severo Ochoa excellence program CEX2019-000920-S and financial support from the Canarian Agency for Research, Innovation and Information Society (ACIISI), of the Canary Islands Government, and the European Regional Development Fund (ERDF), under grant with reference ProID2021010074.

\bmhead{Additional Information}

Correspondence should be addressed to JEM-D: jemd@uni-heidelberg.de








\bibliography{sn-bibliography}


\end{document}